\documentclass[aps,prb,twocolumn,superscriptaddress,amsmath,amssymb,showpacs]{revtex4}

\usepackage{amssymb}
\usepackage[dvips]{graphicx}
\usepackage{bm}
\usepackage{epsfig}
\usepackage[ansinew]{inputenc}

\begin{document}

\title{Probing the localized to itinerant behavior of the 4$f$ electron in CeIn$_{3-x}$Sn$_x$ by Gd$^{3+}$ electron spin resonance}

\author{E. M. Bittar}
\email{eduardo.bittar@lnls.br}
\affiliation{Instituto de F\'{\i}sica ``Gleb Wataghin", UNICAMP, 13083-859, Campinas, SP, Brazil}
\affiliation{Laborat\'{o}rio Nacional de Luz S\'{\i}ncrotron, C. P. 6192, 13083-970, Campinas, SP, Brazil}

\author{C. Adriano}
\affiliation{Instituto de F\'{\i}sica ``Gleb Wataghin", UNICAMP, 13083-859, Campinas, SP, Brazil}

\author{C. Giles}
\affiliation{Instituto de F\'{\i}sica ``Gleb Wataghin", UNICAMP, 13083-859, Campinas, SP, Brazil}

\author{C. Rettori}
\affiliation{Instituto de F\'{\i}sica ``Gleb Wataghin", UNICAMP, 13083-859, Campinas, SP, Brazil}
\affiliation{Centro de Ci\^{e}ncias Naturais e Humanas, Universidade Federal do ABC, 09210-170, Santo Andr\'{e}, SP, Brazil}

\author{Z. Fisk}
\affiliation{Department of Physics and Astronomy, University of California, Irvine, 92697-4575, Irvine, CA, USA}

\author{P. G. Pagliuso}
\affiliation{Instituto de F\'{\i}sica ``Gleb Wataghin", UNICAMP, 13083-859, Campinas, SP, Brazil}
\affiliation{Department of Physics and Astronomy, University of California Irvine, 92697-4575, Irvine, CA, USA}

\date{\today}

\begin{abstract}
The CeIn$_{3-x}$Sn$_x$ cubic heavy fermion system presents an antiferromagnetic transition at $T_N=10$ K, for $x=0$, that decreases continuously down to 0 K upon Sn substitution at a critical concentration of $x_c\approx0.65$. In the vicinity of $T_N\rightarrow0$ the system shows non-Fermi liquid behavior due to antiferromagnetic critical fluctuations. For a high Sn content, $x\gtrsim2.2$, intermediate valence effects are present. In this work we show that Gd$^{3+}$-doped electron spin resonance (ESR) probes a change in the character of the Ce 4$f$ electron, as a function of Sn substitution. The Gd$^{3+}$ ESR results indicate a transition of the Ce 4$f$ spin behavior from localized to itinerant. Near the quantum critical point, on the antiferromagnetic side of the magnetic phase diagram, both localized and itinerant behaviors coexist.
\end{abstract}


\maketitle

\section{Indroduction}

Heavy fermion (HF) systems have shown to the scientific community interesting physical phenomena like antiferromagnetism (AFM), superconductivity (SC), \cite{RevHFStewart} and non-Fermi liquid (NFL) behavior in the vicinity of quantum instabilities. \cite{RevLohneysen} However, the evolution from high-temperature unscreened localized $f$ electrons to itinerant heavy quasiparticles at low temperature is still an open question in condensed matter physics. The description of these HF materials stands on the Kondo lattice model, \cite{RevHFStewart} in which there are three important energy scales: the crystalline electric field (CEF) splitting, the characteristic temperature $T^{*}$, and the single impurity Kondo temperature $T_K$.  The latter is related to the screening of local moments by the conduction electrons due to the Kondo effect, whereas $T^{*}$ represents the crossover between a lattice of Kondo impurities and a coherence state where the hybridization becomes a global process. This energy scale is related to the Ruderman-Kittel–Kasuya–Yosida (RKKY) exchange interaction, since it corresponds to the nearest-neighbor intersite coupling, which is mediated by conduction electrons. \cite{twofluid_pascoal}

The cubic HF CeIn$_{3-x}$Sn$_x$ system is an interesting series for studying the correlations between $T_K$ and $T^{*}$. For $x=0$ the compound is AFM with $T_N=10$ K, and by Sn substitution, $T_N$ decreases continuously down to 0 K at a critical concentration $x_c\approx0.65$. \cite{CeInSn_Lawrence,CeInSn_Euro} This system resembles the behavior of CeIn$_{3}$ under pressure, where an SC state emerges at a critical pressure $P_c\approx25$ kbar with a critical temperature $T_c\approx0.15$ K as $T_N\rightarrow0$. \cite{CeIn3_SC} In the vicinity of $P_c$ and $x_c$ both systems show NFL behavior, suggesting that AFM critical fluctuations are present. Recently, an analysis of the magnetic contribution to the specific heat in CeIn$_{3}$ showed that the magnetic fluctuations in this material are effectively 2D. \cite{CeIn3Cd} Indeed, an almost-linear dependence of $T_N(x)$ is seen for CeIn$_{3-x}$Sn$_x$, \cite{CeInSn_Euro} in contrast to what is predicted by the 3D spin density wave (SDW) theory and it cannot be associated with disorder effects. \cite{CeIn3_Gru} The reported scenario for the pressure and Sn substitution driven quantum critical point (QCP) were different. For CeIn$_{3-x}$Sn$_x$ an SDW description of criticality based on critical exponents analysis of a 3D-AFM was used. \cite{CeIn3Sn025,CeIn3_Gru} In the SDW QCP the 4$f$ moments are delocalized in the AFM state and no change in the Fermi surface is observed across the QCP. \cite{RevLohneysen} However, in a local class of QCP the 4$f$ electrons remain localized in the magnetically ordered phase and there is an abrupt change in the Fermi surface volume at the QCP. \cite{RevLohneysen} For CeIn$_{3}$ under pressure a local QCP was proposed due to a Fermi surface volume change observed in de Haas-van Alphen measurements. \cite{dHvA_CeIn3} Also, some indication of a first-order quantum phase transition instead of a QCP was reported by nuclear quadrupolar resonance measurements carried out around $P_c$. \cite{NQRCep}

In this work we study the evolution of the Gd$^{3+}$ electron spin resonance (ESR) signal in the CeIn$_{3-x}$Sn$_x$ system through its QCP. Since the Ce$^{3+}$ ESR signal is silent, we chose Gd$^{3+}$ as a probe because it is almost a pure $S$-state, so its total angular momentum is mainly due to spin, being weakly perturbed by CEF effects.  To the best of our knowledge no systematic reports on microscopic studies on the Sn substitution $x_c$ were reported. Our Gd$^{3+}$ ESR results show a change in the character of the Ce 4$f$ electron, as a function of Sn substitution, which indicates a transition from localized to itinerant behavior. Near the QCP ($x=0.5$), on the AFM side of the magnetic phase diagram, \cite{CeInSn_Euro,CeIn3_Gru} the Ce 4$f$ spin present simultaneously both localized and itinerant characters.

\section{Experimental details}

Single crystals of Gd doped CeIn$_{3-x}$Sn$_x$ are synthesized by the flux-growth technique. Elemental Ce:Gd:In:Sn are weighted at the ratio 1-$y$:$y$:10-(10$x$/3):10$x$/3, with a nominal value for $y$ of 0.005 and $x$ = 0, 1.5, and 3. Polycrystalline samples are also grown by arc melting in an argon atmosphere. In this case the reactants ratio used is 1-$y$:$y$:3-$x$:$x$, with the same nominal value for $y$ and $x$ = 0, 0.5, 0.7, 1.5, and 3. X-ray powder diffraction measurements confirm the cubic AuCu$_3$ ($Pm$-$3m$)-type structure for all synthesized compounds. The temperature dependence of the magnetic susceptibility, $\chi(T)$, is measured for $2\leq T\leq300$ K, after zero field cooling. All ESR experiments are performed on a fine powder ($d\leq38$~$\mu$m) in a Bruker ELEXSYS $X$-band spectrometer (9.4 GHz) with a TE$_{102}$ cavity coupled to a helium-gas-flux temperature controller system at $4.2\leq T\leq300$ K. Fine powder of crushed single and polycrystals are used in the ESR experiments in order to increase the ESR signal-to-noise ratio. As reference compounds, Gd doped LaIn$_{3-x}$Sn$_x$ alloys were also grown and studied. \cite{ESR_La}

\section{Experimental results}

\begin{figure}[!ht]
\begin{center}
\includegraphics[width=0.5\textwidth]{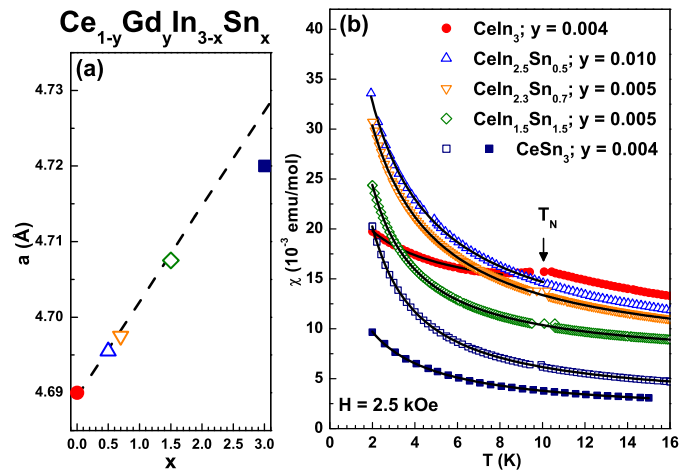}
\vspace{-0.90cm}
\end{center}
\caption{(Gd$^{3+}$ in Ce$_{1-y}$Gd$_y$In$_{3-x}$Sn$_x$. (a) Cubic lattice parameter $a$ dependance as a function of $x$. The dashed line represents the Vegard law. \cite{CeInSn_Euro} For $x=3$ the departure of linear behavior is due to the Ce ions intermediate valence effects for $x\gtrsim2.2$. \cite{CeInSn_Lawrence} (b) Low-temperature dependence of $\chi(T)$ at $H = 2.5$ kOe. Solid lines are the Curie-Weiss fitting. Filled symbols identify single-crystalline samples; open circles, polycrystals.}
\label{chi_xray}
\end{figure}

The actual Sn concentrations are obtained from the cubic lattice parameter, which one expects to follow a linear increase (Vegard's law) \cite{CeInSn_Euro} [see Fig. \ref{chi_xray}(a)]. For $x=3$ the departure of linear behavior is due to the Ce ion intermediate valence effects for $x\gtrsim2.2$. \cite{CeInSn_Lawrence} The temperature dependence of the magnetic susceptibility $\chi(T)$ for the series of compounds Ce$_{1-y}$Gd$_y$In$_{3-x}$Sn$_x$, corrected for the core diamagnetism, is shown in Fig. \ref{chi_xray}(b). From the Curie-Weiss law fitting of the low temperature magnetic susceptibility data, the Gd doping concentration is obtained and its values are listed in Table \ref{Tab_Exp_CeInSn}.

\begin{figure}
\begin{center}
\includegraphics[width=0.5\textwidth]{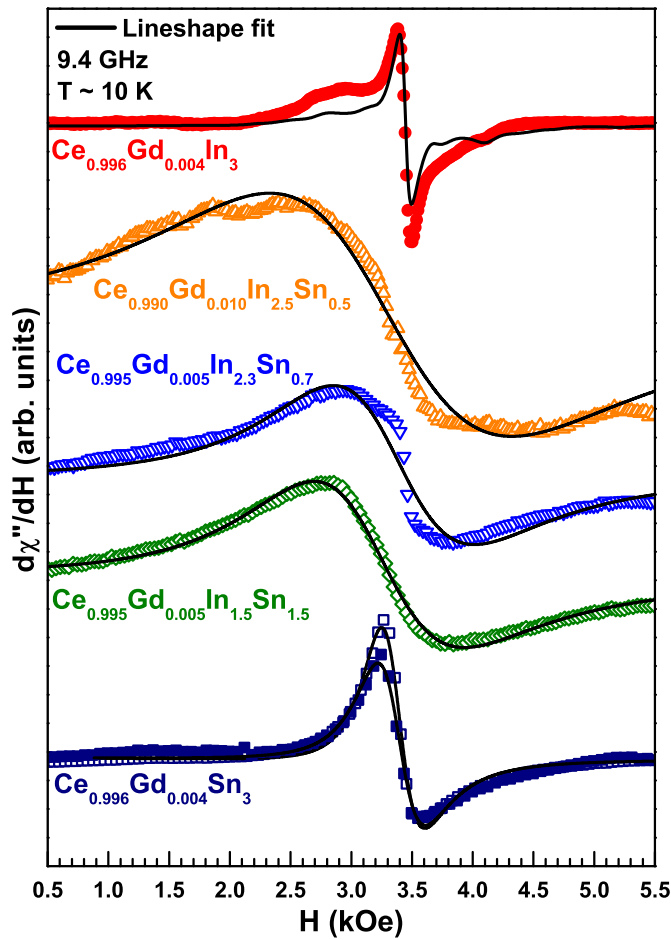}
\end{center}
\caption{Gd$^{3+}$ ESR powder spectra in Ce$_{1-y}$Gd$_y$In$_{3-x}$Sn$_x$ for $y\sim0.5$\%, at $T\approx10$ K, emphasizing the resonance region. Solid lines are the single Dysonian line-shape analysis. For Ce$_{0.996}$Gd$_{0.004}$In$_{3}$ the spin Hamiltonian model discussed in Ref. \onlinecite{Venegas} for powder was used. \cite{ESR_CeIn3} Background contribution is present for the $x=0.5$ and $x=0.7$ samples (see text). Filled symbols identify single-crystalline samples; open circles, polycrystals.}
\label{esr}
\end{figure}

Figure \ref{esr} shows the ESR ($X$-band) powder spectra, at $T\sim10$ K, of Gd$^{3+}$ in CeIn$_{3-x}$Sn$_x$. Except for $x=0$, the ESR spectra consist of a single Dysonian resonance, consistent with the ESR for localized magnetic moments in a metallic host with a skin depth smaller than the size of the used particles. By fitting the line shape to the appropriate admixture of absorption and dispersion Lorentzian derivatives, we obtain the $g$ value and line width $\Delta H$ of the resonances. The solid lines are the best fit to the observed resonances and the obtained $g$ shifts $\Delta g$ [relative to the $g=1.993(1)$ seen in cubic insulators] are presented in Table \ref{Tab_Exp_CeInSn}. For Gd$^{3+}$ in CeIn$_{3}$ the ESR spectrum shows the typical fine-structure features for powder samples, \cite{gambke} with a main line at $H\sim3.45$ kOe, associated with the $1/2\leftrightarrow1/2$ transition. A previous report on this compound, using the spin Hamiltonian $H=g\mu_{B}\textbf{H}\cdot\textbf{S}+(1/60)b_4(O_{4}^{0}+5O_{4}^{4})+J_{fs}\textbf{S}\cdot \textbf{s}$, \cite{Venegas} extracted the crystal field parameter $b_4=90(5)$ Oe. \cite{ESR_CeIn3} For Ce$_{0.995}$Gd$_{0.005}$In$_{2.3}$Sn$_{0.7}$ a background line is present in the spectrum at $H\sim3.4$ kOe and for Ce$_{0.99}$Gd$_{0.01}$In$_{2.5}$Sn$_{0.5}$ the ESR spectrum also shows some small contribution of the background. These background contributions are due to extrinsic impurities, with a resonance at $g\sim2$, present in the cavity or in the cryostat quartz (even without any sample).

\begin{figure}
\begin{center}
\includegraphics[width=0.5\textwidth]{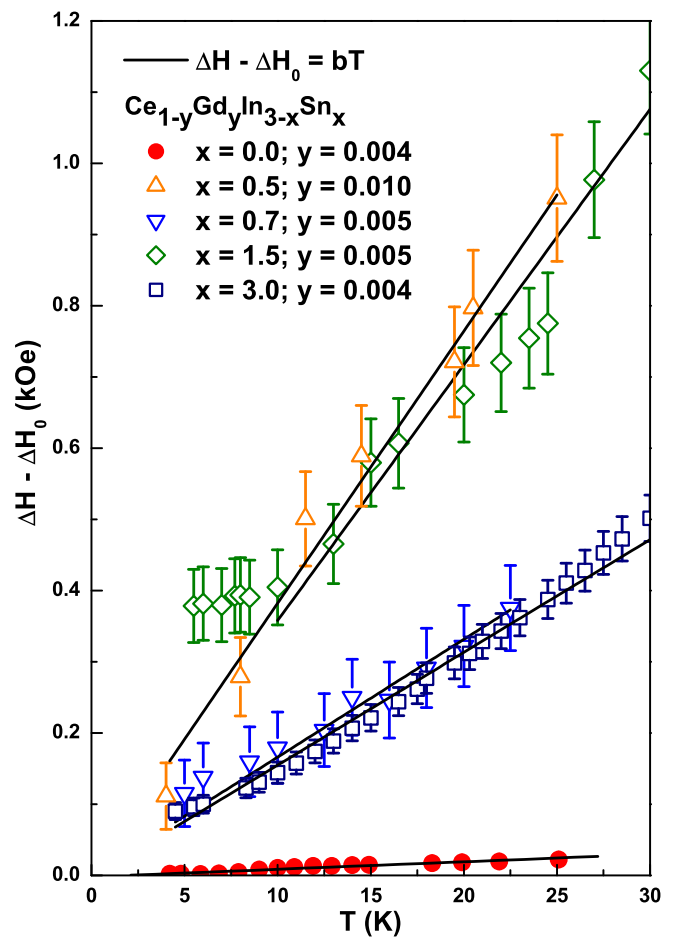}
\end{center}
\caption{Temperature dependence of the Gd$^{3+}$ ESR linewidth in Ce$_{1-y}$Gd$_y$In$_{3-x}$Sn$_x$. Solid lines are the best fit to $\Delta H-\Delta H_0=bT$. A deviation from the linear dependence of $\Delta H$ at low temperatures is seen for $x=1.5$, which is related to short-range Gd-Gd interaction. Filled symbols identify single-crystalline samples; open circles, polycrystals.}
\label{korringa}
\end{figure}

\begin{figure}
\begin{center}
\includegraphics[width=0.5\textwidth]{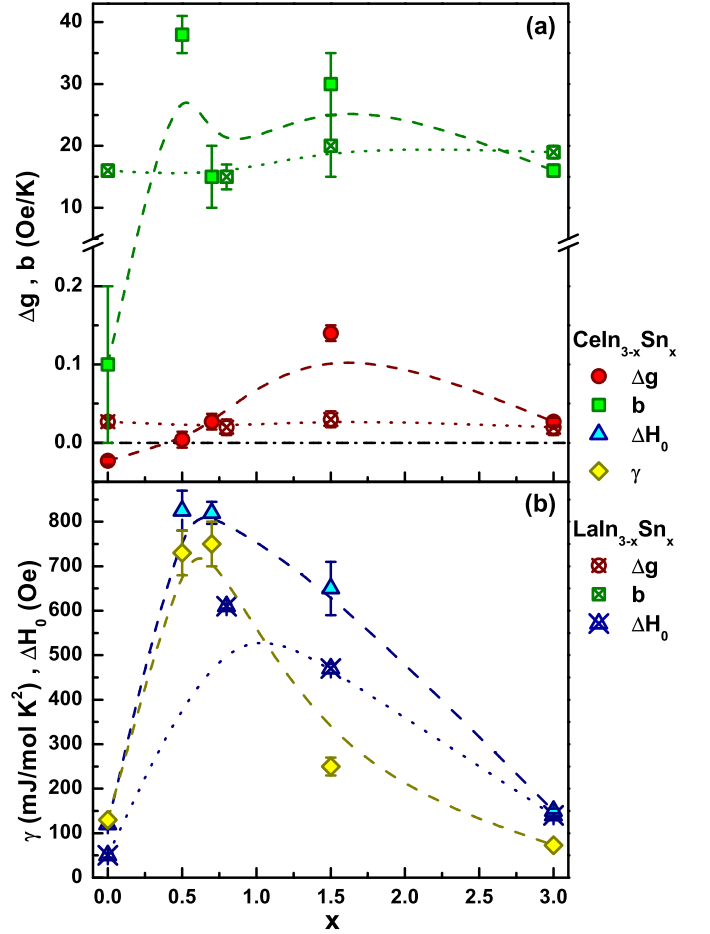}
\end{center}
\caption{Gd$^{3+}$ ESR in Ce$_{1-y}$Gd$_y$In$_{3-x}$Sn$_x$. (a) Line-width thermal broadening $b$ and $g$ shift evolution. The dash-dotted line marks $\Delta g=0$. (b) Sommerfeld coefficient $\gamma$ and residual line-width $\Delta H_0$ evolution. For comparison, the Gd$^{3+}$ ESR in La$_{1-y}$Gd$_y$In$_{3-x}$Sn$_x$ data are also shown. \cite{ESR_La} Dashed and dotted spline lines are guides for the eye.}
\label{Fig_CeInSn_Exp}
\end{figure}

The temperature dependence of $\Delta H$ is shown in Fig. \ref{korringa}. For all samples there is a range where the width increases linearly with temperature. In this range the linear dependence of the $\Delta H$ is fitted to the expression $\Delta H-\Delta H_0=bT$. The values for $\Delta H_0$ (residual line width) and $b$ (line-width thermal broadening) are presented in Table \ref{Tab_Exp_CeInSn} and Fig. \ref{Fig_CeInSn_Exp}. The relatively high $\Delta H_0$ values for $0<x<3$ are probably due to unresolved CEF and disorder introduced by the In-Sn substitution. The $\Delta H_0$ values follow the residual electrical resistivity $\rho_0$ behavior, since both are dependent on the disorder. One can see that $\Delta H_0$ has the same pattern for LaIn$_{3-x}$Sn$_x$ [Fig. \ref{Fig_CeInSn_Exp}(b)], which follows the $\rho_0$ dependence (see Fig. 4 in Ref. \onlinecite{LaInSnrho}). For Ce$_{0.996}$Gd$_{0.004}$In$_{3}$, only the $\Delta H$ temperature dependance of the main line is analyzed. A deviation from the linear dependence of $\Delta H$ at low temperature for $x=1.5$ is related to short-range Gd-Gd interaction. Within the accuracy of the measurements, the $g$ and $b$ values are Gd concentration independent for $y<1.0\%$ (not shown). Therefore, bottleneck and dynamic effects can be disregarded. \cite{ref20}

Since in this work samples with different Sn content are grown by different methods, we show, particularly for $x=3$ (Fig. \ref{esr}), that there are no discernible differences in the ESR spectra of grounded single- vs polycrystalline samples. Hence, for the CeIn$_{3-x}$Sn$_x$ system, single and polycrystals are indistinguishable from the ESR point of view. However, this is not always the case in most systems and it cannot be established \textit{a priori}. Table \ref{Tab_Exp_CeInSn} and Fig. \ref{Fig_CeInSn_Exp} do not distinguish single and polycrystalline samples.

\begin{table}
\caption{Experimental parameters for Gd$^{3+}$ diluted in Ce$_{1-y}$Gd$_y$In$_{3-x}$Sn$_x$. Values of $\gamma$ are taken from Ref. \onlinecite{CeInSn_Euro}.} \label{Tab_Exp_CeInSn} \centering
\begin{tabular}{cccccc}
\hline \hline
Gd$^{3+}$ & Gd &  & $\Delta H_0$ & $b$ & $\gamma$ \\
in & $y$
 & $\Delta g$ & [Oe] & [Oe/K] & [mJ/mol K$^{2}$] \\
\hline
CeIn$_{3}$ & 0.004 & -0.023(5) & 120(5) & 0.1(1) & 130 \\
CeIn$_{2.5}$Sn$_{0.5}$ & 0.010 & +0.007(10) & 825(45) & 38(3) & 730(50) \\
CeIn$_{2.3}$Sn$_{0.7}$ & 0.005 & +0.027(10) & 820(25) & 15(5) & 750(50) \\
CeIn$_{1.5}$Sn$_{1.5}$ & 0.005 & +0.140(10) & 650(60) & 30(5) & 250(20) \\
CeSn$_{3}$ & 0.004 & +0.027(5) & 150(5) & 16(1) & 73 \\
\hline \hline
\end{tabular}
\end{table}

\begin{figure}
\begin{center}
\includegraphics[width=0.5\textwidth]{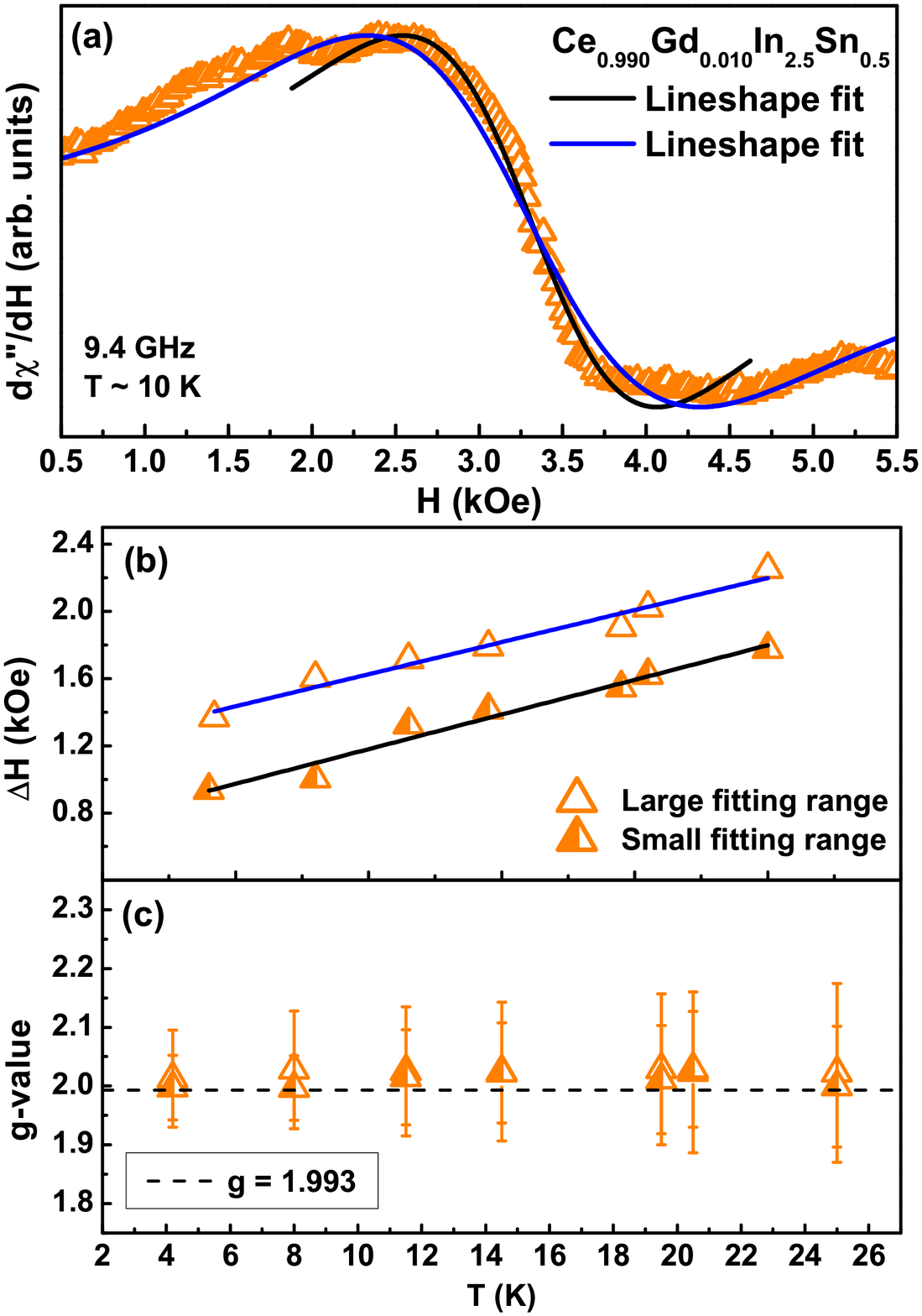}
\end{center}
\caption{(a) Gd$^{3+}$ ESR powder spectra in Ce$_{0.990}$Gd$_{0.010}$In$_{2.5}$Sn$_{0.5}$, at $T\approx10$ K. Solid lines are the single Dysonian line-shape analysis for two field range fittings. (b) Temperature dependence of the Gd$^{3+}$ ESR linewidth for the two field ranges shown in (a). Solid lines are the best fit to $\Delta H=\Delta H_0+bT$. (c) $g$-value temperature dependence of the Gd$^{3+}$ ESR for the two field ranges shown in (a). The dashed line is the $g$ value for Gd$^{3+}$ ESR in insulators.}
\label{x05}
\end{figure}

\begin{figure}
\begin{center}
\includegraphics[width=0.5\textwidth]{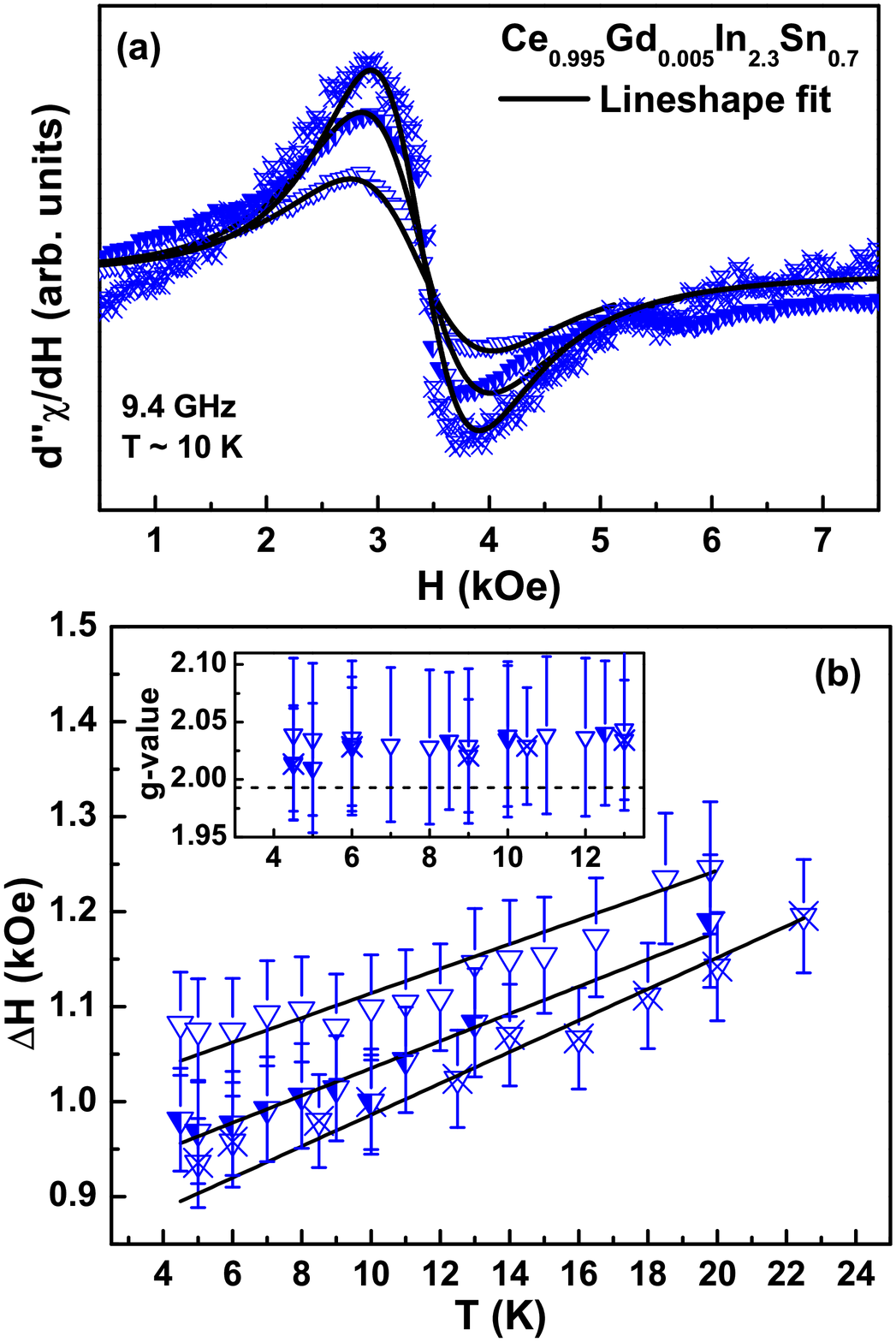}
\end{center}
\caption{(a) Gd$^{3+}$ ESR powder spectra in Ce$_{0.995}$Gd$_{0.005}$In$_{2.3}$Sn$_{0.7}$, at $T\approx10$ K, for three samples with the same Sn concentration. Solid lines are the single Dysonian line-shape analysis. (b) Temperature dependence of the Gd$^{3+}$ ESR linewidth for the three samples shown in (a). Solid lines are the best fit to $\Delta H=\Delta H_0+bT$. \textit{Inset}: Low-temperature $g$-value dependence of the Gd$^{3+}$ ESR for the three different samples shown in (a). The dashed line is the $g$ value for Gd$^{3+}$ ESR in insulators.}
\label{x07}
\end{figure}

The error bar values presented in Table \ref{Tab_Exp_CeInSn} and Fig. \ref{Fig_CeInSn_Exp} are determined by systematic measurements of different samples for most Sn concentrations and by analyzing the line-shape fitting for different field ranges. To exemplify such systematic procedures, Fig. \ref{x05} illustrates a line-shape analysis for different field range fittings, in this case for the Ce$_{0.990}$Gd$_{0.010}$In$_{2.5}$Sn$_{0.5}$ sample. One can observe that the slope of the line-width thermal broadening [Fig. \ref{x05}(b)] and the $g$-value [Fig. \ref{x05}(c)] of the Gd$^{3+}$ ESR are almost independent of the field range fitting. Also, Fig. \ref{x07} exemplifies different samples measurements for Ce$_{0.995}$Gd$_{0.005}$In$_{2.3}$Sn$_{0.7}$. Again, the $b$ and $g$  values [Fig. \ref{x07}(b)] vary little between samples. Therefore, despite the large line width of the Gd$^{3+}$ ESR in Ce$_{1-y}$Gd$_y$In$_{3-x}$Sn$_x$, which would give rise to large error values, our systematic measurements and fitting procedures allow us to reduce the error and determine the values with a higher precision.

\section{Analysis}

\subsection{Gd$^{3+}$ ESR in metals}

In metals, the exchange interaction $J_{fs}(q)\mathbf{S}\cdot\mathbf{s}$ between a Gd$^{3+}$ localized 4$f$ electron spin ($\mathbf{S}$) and the conduction electron spin ($\mathbf{s}$) of the host metal yields an ESR $\Delta g$ (Knight shift) given by \cite{EPR_barnes}
\begin{equation}
\Delta g=J_{fs}(0)\eta _{F_s}, \label{Eq1}
\end{equation}
where $J$$_{fs}(0)$ is the effective exchange interaction parameter between the Gd$^{3+}$ 4$f$ local moment and the $s$-like conduction electrons in the absence of conduction electron momentum transfer $(q=|\mathbf{k}-\mathbf{k}^{\prime}|=k_F[2(1-cos\theta_{\mathbf{k}\mathbf{k}^{\prime}})]^{1/2}=0)$. \cite{DavidovSSC12} $\eta_{F_s}$ is the $s$-like-band bare density of states for one spin direction at the Fermi surface.

In addition, the exchange interaction leads to a thermal broadening of $\Delta H$, $b$ (Korringa rate), given by \cite{EPR_barnes}
\begin{equation}
b=\frac{d(\Delta H)}{dT}=\frac{\pi k_{B}}{g\mu _{B}}J_{fs}(0)^{2}\eta _{F_s}^{2},  \label{Eq2}
\end{equation}
where the constants $k_B$, $\mu_B$ and $g$ are the Boltzman constant, the Bohr magneton, and the Gd$^{3+}$ $g$ value in insulators ($g=1.993$), respectively. The constant $\pi k_{B}/g\mu _{B}$ is $2.34\times10^{4}$~Oe/K in CGS units.

Equations \ref{Eq1} and \ref{Eq2} are normally used in the analysis of ESR data for noninteracting and highly diluted rare-earths magnetic moments in intermetallic compounds with appreciable residual resistivity, i.e., large conduction electrons spin-flip scattering (absence of ``bottleneck'' and ``dynamic'' effects). \cite{ref20} Combining the above equations we can write
\begin{equation}
b=\frac{\pi k_{B}}{g\mu _{B}}(\Delta g)^{2}. \label{Eq3}
\end{equation}

When the effective exchange interaction constant is not independent of the momentum transfer $(q\neq0)$, Eq. \ref{Eq2}, in this more general case, has to be rewritten as
\begin{equation}
b=\frac{\pi k_{B}}{g\mu _{B}}\left\langle J_{fs}^{2}(q)\right\rangle\eta _{F_s}^{2}, \label{Eq4}
\end{equation}
or alternatively, using Eq. \ref{Eq1}
\begin{equation}
b=\frac{\pi k_{B}}{g\mu _{B}}\frac{\left\langle J_{fs}^{2}(q)\right\rangle_F}{J_{fs}^2(0)}\Delta g ^2, \label{Eq5}
\end{equation}
where $\langle J_{fs}^2(q)\rangle_{F}$ is the square of the effective exchange interaction parameter in the presence of conduction electron momentum transfer, averaged over the Fermi surface. \cite{DavidovSSC12}

One way to know if the system is momentum transfer dependent is to analyze Eq. \ref{Eq3}. If the calculated Korringa rate $b_{cal}$ by the experimental $\Delta g$ is equal to the experimental Korringa rate $b_{exp}$ ($b_{cal}=b_{exp}$), $q$ dependance can be neglected. However, if $b_{cal}>b_{exp}$ then it cannot. This is because $\langle J_{fs}^{2}(q)\rangle_F/J_{fs}^2(0) \leq 1$, once $J(q)$ is proportional the Fourier transformation of $J(r)$, which amplitude decreases as function of $r$. So the average $J(r)$ should be smaller than $J(0)$.

In cases where the conduction band has also $d$-, $p$- or $f$-like electrons, Eqs.~\ref{Eq1} and \ref{Eq2} are not valid and must be rewritten, respectively as
\begin{eqnarray}
\Delta g & = & \Delta g_{fs}+\Delta g_{fd}+\Delta g_{fp}+... \nonumber \\
 & = & J_{fs}(0)\eta _{F_s}+J_{fd}(0)\eta _{F_d}+J_{fp}(0)\eta _{F_p}+... \label{Eq6}
\end{eqnarray}
and
\begin{eqnarray}
b & = & \frac{\pi k_{B}}{g\mu _{B}}[F_s J_{fs}^{2}(0)\eta _{F_s}^{2}+ \nonumber \\
  & & +F_d J_{fd}^{2}(0)\eta _{F_d}^{2}+F_p J_{fp}^{2}(0)\eta _{F_p}^{2}+...] \label{Eq7}
\end{eqnarray}
where $J_{fs}(0)$, $J_{fd}(0)$, and $J_{fp}(0)$ are the exchange interaction constants between the Gd$^{3+}$ 4$f$ spin and the $s$-, $d$-, and $p$-like bands, respectively. $\eta _{F_s}$, $\eta _{F_d}$, and $\eta _{F_p}$ are the bare density of states for one spin direction at the Fermi surface for each respective band. $F_s=1$, $F_d=1/5$, and $F_p=1/3$ are factors associated with the orbital degeneracy of the unsplit (no CEF effects) $s$, $d$, and $p$ bands at the Fermi level, respectively. \cite{Gaston,Yafet}

Multiband effects enhance the Korringa rate compared to $b_{cal}$ (Eq. \ref{Eq3}), since the dependence of $b$ is quadratic with the exchange interaction parameters, while for $\Delta g$ it is linear and depends on the sign and strength of each exchange interaction constant. Therefore, the $\Delta g$ sign can give valuable information about the interaction between the localized moment and its environment.

\subsection{G\lowercase{d}$^{3+}$ effective exchange interaction parameter calculations in C\lowercase{e}I\lowercase{n}$_{3-x}$S\lowercase{n}$_x$}

We now analyze separately the experimental ESR data for each synthesized compound.

\subsubsection{Calculation for $x=0.0$}

In the absence of strong electron-electron exchange interaction and assuming that $\langle J_{fs}^2(q)\rangle_{F}^{1/2}=J_{fs}(0)$, i.e., the effective exchange interaction is constant over the Fermi surface, one expects $b\approx12(5)$ Oe/K from Eq. \ref{Eq3}, using the experimental $\Delta g\simeq-23(5)\times 10^{-3}$. This value is much larger than that measured experimentally, $b=0.1(1)$ Oe/K (Fig. \ref{korringa}). Thus, the approximations that the relaxation does not depend on $q$ and that it is due to the contribution of a single conduction $s$-like band are not adequate. Since $\Delta g$ is negative, a relaxation via a single $s$ band is not plausible because $J_{fs}(0)$ is atomic-like and positive. Thus, for $\Delta g<0$, contributions coming from covalent-like (negative) exchange interaction between the Gd$^{3+}$ 4$f$-electron and $p$ or $f$ bands must be taken into account in the relaxation process (multiband effects). \cite{EPR_barnes} On the other hand, multiple bands would lead to a Korringa rate higher than the one expected from the $\Delta g$, \cite{ESRYbAl3} contrary to what is observed for Gd$^{3+}$ in CeIn$_3$. Therefore, a strong $q$-dependent effective exchange interaction parameter $J_{fp}(q)$ or $J_{ff}(q)$ is expected in this compound. For CeIn$_{3}$ the local magnetic moment of Ce is compensated by the conduction electron sea due to the Kondo effect. However, when Gd$^{3+}$ substitutes the Ce ions there is a strong Coulomb repulsion potential that decreases the local density of states at the Gd$^{3+}$ site, hence decreasing the Korringa rate [Eq. \ref{Eq2} or \ref{Eq4}]. Theoretical calculations have already shown that the spin relaxation rate of a well-defined magnetic moment in the neighborhood of a fluctuating valence ion decreases in relation to the relaxation rate of an undoped metal. \cite{Pinto} Indeed, a much higher Korringa rate $b=16(1)$ Oe/K was measured in Gd-doped LaIn$_3$. \cite{ESR_CeIn3} This has also been observed for Gd in CePd$_3$ which presented an ESR $\Delta H$ thermal broadening five times smaller than in LaPd$_3$. \cite{ESR_Pd3} Besides, the observation of fine-structure features in the spectrum (Fig. \ref{esr}) even up to room temperature without narrowing effects \cite{EPR_barnes} (not shown) suggests a low local density of states at the Gd$^{3+}$ site. Another consequence of the screening of the Ce$^{3+}$ magnetic moment by conduction electrons is that the Gd$^{3+}$ resonance does not sense the internal field caused by the AFM transition. No change in the relaxation or in the resonance field is observed below $T_N=10$ K.

From the considerations above we can assume that the interaction of the Gd$^{3+}$ 4$f$ local moment is mainly with Ce $f$-like conduction electrons. We then can rewrite Eqs.~\ref{Eq1} and \ref{Eq4}, respectively, as
\begin{equation}
\Delta g=J_{ff}(0)\eta _{F_f}  \label{eqgff}
\end{equation}
and
\begin{equation}
b=\frac{\pi k_{B}}{g\mu _{B}}F_f\left\langle J_{ff}^2(q)\right\rangle_{F}\eta _{F_f}^{2}, \label{eqbff}
\end{equation}
where $F_f=1/7$ is associated with the orbital degeneracy of the unsplit $f$ band at the Fermi level. \cite{Gaston,Yafet}

In the free conduction electron gas model, the electronic heat capacity or Sommerfeld coefficient $\gamma$ is given by
\begin{equation}
\gamma=(2/3)\pi ^{2}k_{B}^{2}\eta _{F} \label{gamma}
\end{equation}
and one can obtain, using its experimental value, the bare density of states for one spin direction at the Fermi surface.

For CeIn$_{3}$ $\gamma^{x=0}=130$ mJ/(mol K$^2$), \cite{CeInSn_Euro} so we get from Eq. \ref{gamma} $\eta _{F}^{x=0}=28(2)$ states/(eV mol spin). Assuming that in this compound the density of states at the Fermi level for the 4$f$ electrons $\eta _{F_f}^{x=0}$ is
\begin{equation}
\eta _{F_f}^{x=0}=\eta _{F}^{x=0}-\eta _{F}^{LaIn_3}, \nonumber
\end{equation}
where $\eta _{F}^{LaIn_3}=0.8(1)$ states/(eV mol spin) (see Fig. 4 in Ref. \onlinecite{LaInSnSC}), we calculate $\eta _{F_f}^{x=0}=27(2)$ states/(eV mol spin).

Using Eqs. \ref{eqgff} and \ref{eqbff}, experimental values of $\Delta g$ and $b$, and $\eta _{F_f}^{x=0}=27(2)$ states/(eV mol spin), we obtain $J_{ff}(0)=-0.8(1)$ meV  and $\langle J_{ff}^{2}(q)\rangle_{F}^{1/2}=0.20(5)$ meV.

\subsubsection{Calculation for $x=0.5$}

By substituting 16.67\% of In by Sn, $x=0.5$, $T_N$ drops to $\sim1.3$ K, \cite{CeInSn_Euro} very close to $x_c$. We also see $\Delta g$ going from a relatively large negative to a very small, $\simeq7(10)\times 10^{-3}$ positive value. From Eq. \ref{Eq3} we get $b_{cal}\ll b_{exp}$. It is clear that for Ce$_{0.990}$Gd$_{0.010}$In$_{2.50}$Sn$_{0.50}$ multiband effects are now present. \cite{ESRYbAl3} This is expected since Sn substitution lead to the hybridization of the localized Ce$^{3+}$ 4$f$ electrons, turning them into an itinerant $s$-like conduction band. So, for $x=0.5$ the Gd$^{3+}$ resonance relaxes via the contribution of the Ce 4$f$ itinerant $s$- and localized $f$-like bands. In this case, Eqs. \ref{Eq6} and \ref{Eq7} can be rewritten, respectively, as
\begin{equation}
\Delta g=J_{fs}\eta _{F_f^{it}}+J_{ff}\eta _{F_f^{loc}} \label{Eq11}
\end{equation}
and
\begin{equation}
b=\frac{\pi k_{B}}{g\mu _{B}}\left[J_{fs}^2\eta _{F_f^{it}}^{2}+F_f J_{ff}^2\eta _{F_f^{loc}}^{2}\right], \label{Eq12}
\end{equation}
where $\eta _{F_f^{it}}$ and $\eta _{F_f^{loc}}$ are the band bare density of states for one spin direction at the Fermi surface for the itinerant- and localized-like 4$f$ band, respectively.

From $\gamma^{x=0.5}=730(50)$ mJ/(mol K$^2$) \cite{CeInSn_Euro} and Eq. \ref{gamma}, we get $\eta _{F}^{x=0.5}=155(2)$ states/(eV mol spin). Assuming that
\begin{equation}
\eta _{F_f}^{x=0.5}=\eta _{F}^{x=0.5}-\eta _{F}^{LaIn_{2.5}Sn_{0.5}}, \nonumber
\end{equation}
where $\eta _{F}^{LaIn_{2.5}Sn_{0.5}}=0.8(1)$ states/(eV mol spin) (see Fig. 4 in Ref.~\onlinecite{LaInSnSC}), we calculate $\eta _{F_f}^{x=0.5}=154(2)$ states/(eV mol spin).

Solving the system of three equations below for $\eta _{F_f}^{x=0.5}=154(2)$ states/(eV mol spin), $\Delta g\simeq7(10)\times 10^{-3}$, $b=38(3)$ Oe/K, and admitting that $J_{ff}^{x=0.5}=J_{ff}^{x=0}=0.0008$ eV,
\begin{equation}
\eta _{F_f}^{x=0.5}=\eta _{F_f^{it}}+\eta _{F_f^{loc}}=154, \nonumber
\end{equation}
\begin{equation}
\Delta g=J_{fs}\eta _{F_f^{it}}+0.0008\eta _{F_f^{loc}}=7\times 10^{-3} \nonumber
\end{equation}
and
\begin{equation}
b=2.34\times10^{4}\left[J_{fs}^2\eta _{F_f^{it}}^{2}+\frac{1}{7}(0.0008)^2\eta _{F_f^{loc}}^{2}\right]=38,  \nonumber
\end{equation}
we obtain $J_{fs}(0)=0.3(1)$ meV, $\eta _{F_f^{it}}=115(10)$ states/(eV mol spin), and $\eta _{F_f^{loc}}=40(5)$ states/(eV mol spin).

So, naively, this result indicates that a weight of 74\% of the Ce $f$-electrons becomes itinerant upon $x=0.5$ Sn substitution while the other 26\% remains localized. One may argue that the multiband effects would be in fact due to the presence of $s$ electrons arising from a weakened Kondo interaction at the Ce$^{3+}$ site or by the addition of new electrons. However, small Sn substitution increases the conduction electrons attractive potential \cite{Pinto} and does not profoundly change the density of states, as seen in LaIn$_{3-x}$Sn$_x$, \cite{ESR_La,LaInSnSC} favoring the interpretation of a delocalization of the Ce $f$-electrons.

\subsubsection{Calculation for $x=0.7$}

For Gd$^{3+}$ in CeIn$_{2.3}$Sn$_{0.7}$ the system is in the vicinity of the QCP and Eq. \ref{Eq3} predicts $b_{cal}\approx b_{exp}$. Therefore, we can consider a single $s$-like conduction band with no $q$ dependance in the analysis of the resonance in this material. Hence, from $\gamma^{x=0.7}=750(50)$ mJ/(mol K$^2$) \cite{CeInSn_Euro} and Eq. \ref{gamma}, we get $\eta _{F}^{x=0.7}=160(10)$ states/(eV mol spin). Using Eq. \ref{Eq2} we find $J_{fs}(0)=0.2(1)$ meV, similar to the value found for $x=0.5$.

\subsubsection{Calculation for $x=1.5$}

The $\Delta g$ value observed experimentally gives $b_{cal}\gg b_{exp}$ by Eq. \ref{Eq3}. So, in this case $q$ dependence is present and $\langle J_{fs}^2(q)\rangle_{F}^{1/2}\neq J_{fs}(0)$. From $\gamma^{x=1.5}=250(20)$ mJ/(mol K$^2$) \cite{CeInSn_Euro} and Eq. \ref{gamma} we get $\eta _{F}^{x=1.5}=53(4)$ states/(eV mol spin). Using Eqs. \ref{Eq1} and \ref{Eq4} we calculate $J_{fs}(0)=2.6(2)$ meV and $\langle J_{fs}^2(q)\rangle_{F}^{1/2}=0.7(1)$ meV, respectively.

\subsubsection{Calculation for $x=3.0$}

From Eq. \ref{Eq3} we get $b_{cal}\approx b_{exp}$, i.e., multiband and $q$ dependence effects of the exchange interaction may be neglected. Thus, from $\gamma^{x=3}=73$ mJ/(mol K$^2$) \cite{CeInSn_Euro} and Eq. \ref{gamma} we get $\eta _{F}^{x=3}=16(1)$ states/(eV mol spin). Using Eq. \ref{Eq2} we find $J_{fs}(0)=1.7(1)$ meV.

\subsection{Derived effective exchange interaction parameters summary}

The derived effective exchange interaction parameters from the analysis above are summarized in Table \ref{Tab_Cal_CeInSn}. Due to the suppositions and approximations considered in the calculations, the numerical values must be taken with care. However, it does not invalidate the qualitative microscopic description probed by ESR.

\begin{table}[htb!]
\caption{Derived effective exchange interaction parameters for Gd$^{3+}$ diluted in CeIn$_{3-x}$Sn$_x$.} \label{Tab_Cal_CeInSn}\centering
\begin{tabular}{ccccc}
\hline
\hline
Gd$^{3+}$ & $J_{fs}(0)$ & $\left\langle J_{fs}^2(q)\right\rangle_{F}^{1/2}$ & $\left|J_{ff}(0)\right|$ & $\left\langle J_{ff}^2(q)\right\rangle_{F}^{1/2}$ \\
in & [meV] & [meV] & [meV] & [meV] \\
 \hline
CeIn$_{3}$ & & & 0.8(1) & 0.20(5)\\
CeIn$_{2.5}$Sn$_{0.5}$ & 0.3(1) & & 0.8(1) & \\
CeIn$_{2.3}$Sn$_{0.7}$ & 0.2(1) & & & \\
CeIn$_{1.5}$Sn$_{1.5}$ & 2.6(2) & 0.7(1) & & \\
CeSn$_{3}$ & 1.7(1) & & & \\
\hline
\hline
\end{tabular}
\end{table}

\section{Discussion}

\begin{figure}
\begin{center}
\includegraphics[width=0.5\textwidth]{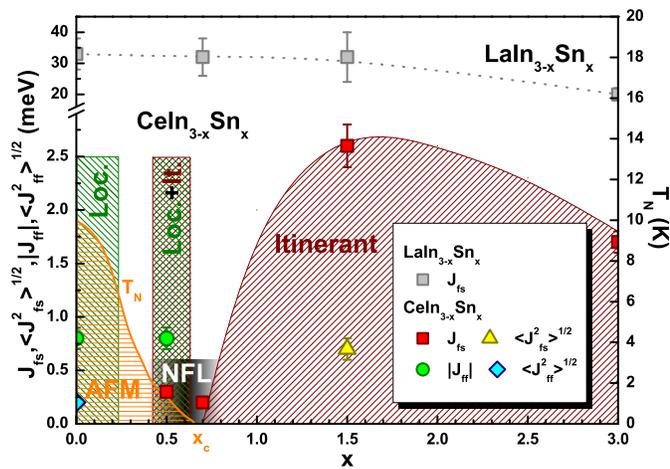}
\end{center}
\caption{Effective exchange interaction parameter evolution as a function of Sn substitution. Data for LaIn$_{3-x}$Sn$_x$ compounds are taken from Ref. \onlinecite{ESR_La}. The AFM temperature transition $T_N$ evolution and the non-Fermi liquid (NFL) region in the vicinity of the critical Sn concentration $x_c$ are also shown. \cite{CeInSn_Euro} For CeIn$_{3-x}$Sn$_x$ one can identify the presence of only the localized (Loc.) spin behavior for $x=0$ and the itinerant (It.) character for $x\geq0.7$, since only a single band effective exchange interaction is probed in each case. At $x=0.5$, near the quantum critical point, Gd$^{3+}$ ESR probes both localized and itinerant components of the Ce 4$f$ electron. Shaded areas and spline lines are guides for the eye.}
\label{Fig_CeInSn_Phase}
\end{figure}

The nonmagnetic analog LaIn$_{3-x}$Sn$_x$ system is superconducting and shows Pauli paramagnetism in the normal sate. \cite{LaInSnSC} Gd$^{3+}$-doped ESR measurements in these compounds showed that the $g$ shift and Korringa rate are not strongly changed by Sn substitution [see Fig. \ref{Fig_CeInSn_Exp}(a)]. \cite{ESR_La} The Gd$^{3+}$ relaxation in these alloys is always via a single $s$-like conduction band and $J_{fs}$ is $q$ independent, decreasing slightly with increasing $x$  (Fig. \ref{Fig_CeInSn_Phase}). \cite{ESR_La}

For CeIn$_{3-x}$Sn$_x$ compounds the evolution of the Gd$^{3+}$ ESR with Sn substitution is not as straightforward as in LaIn$_{3-x}$Sn$_x$.  The $b$ and $\Delta g$ values change profoundly as a function of $x$ [Fig. \ref{Fig_CeInSn_Exp}(a)]. For CeIn$_3$, as we have seen, there is no exchange interaction between Gd$^{3+}$ and the $s$-like conduction electrons. This is due to the Kondo effect, which creates an attractive potential for these $s$-like conduction electrons at the Ce sites, reducing its density at the Gd site. In this case, the Gd$^{3+}$ ESR relaxes only via an $f$-like localized band. This attests that the Ce 4$f$ electrons in CeIn$_3$ are strongly localized under high Kondo screening, which also prevents the Gd$^{3+}$ resonance from sensing the AFM transition below $T_N=10$ K.

As the system approaches the QCP ($x=0.5$), but still presenting AFM order, we observe the appearance of multiband effects in the resonance which are related to the delocalization of the 4$f$ electrons, induced by the Sn substitution, giving rise to an $s$-like band. For this alloy the Ce 4$f$ electrons coexist as localized and itinerant. In the vicinity of the QCP ($x=0.7$), on the nonmagnetic side of the phase diagram, the resonance assumes a character where the relaxation is via a single $s$-like band. The effective exchange interaction parameter $J_{fs}(0)$ in this compound is, within experimental errors, the same as at $x=0.5$, but no local $f$-like electrons are probed by the Gd$^{3+}$; only the itinerants.

Further increase in the Sn substitution does not alter the Gd$^{3+}$ relaxation process, which remains being via a single $s$-like conduction band. For $x=1.5$ the exchange interaction is $q$ dependent, indicating that it is not constant over the Fermi surface and this dependance might be related to an anisotropy observed in the $s$-$f$ hybridization for CeIn$_1$Sn$_2$. \cite{Ani_CeIn1Sn2} In CeSn$_3$ the $J_{fs}(0)$ value decreases slightly compared to CeIn$_{1.5}$Sn$_{1.5}$, probably due to intermediate valence effects and/or lattice expansion. Thus, once the system crosses the QCP the hybridization of the localized 4$f$ electrons with the conduction band becomes a global process and it behaves only as an itinerant.

Figure \ref{Fig_CeInSn_Phase} qualitatively summarizes this discussion of Gd$^{3+}$ ESR evolution in CeIn$_{3-x}$Sn$_x$ materials.

\section{Conclusions}

For CeIn$_{3-x}$Sn$_x$, our conclusions are not drawn solely based on the values of the extracted or derived parameters, as a function of Sn concentration, but mostly based on the fact that one cannot analyze in the same way the Gd$^{3+}$ ESR of each sample. When comparing the LaIn$_{3-x}$Sn$_x$\cite{ESR_La} and CeIn$_{3-x}$Sn$_x$ systems, one can immediately realize that the evolution upon substituting In by Sn is dramatically different. While for La-based compounds only a slight change in the $J_{fs}(0)$ value is observed, for the HF one, the Gd$^{3+}$ effective exchange parameter alters significantly, depending on the $x$ value. However, the only difference between these systems is the addition of an 4$f$ electron. So, the discrepancy in the evolutionary behavior must come from the physics of this extra 4$f$ electron. For the $x=0$ end member this additional electron is localized and highly screened by the conduction electron sea, and thus, Gd$^{3+}$ ESR only probes an $f$-like localized band. However, for the other end member, $x=3$, no local magnetism occurs and the compound can be described as an HF Landau Fermi liquid, where the Gd$^{3+}$ resonance relaxes only via a single $s$-like itinerant band. On the other hand, in between, specifically for $x=0.5$, we observe multiband effects on the ESR data, i.e., contributions of localized and itinerant bands that are originated from the same Ce extra 4$f$ electron. Therefore, we argue that the microscopic evolution of the 4$f$ electron in the CeIn$_{3-x}$Sn$_x$ system, as a function of Sn substitution, can be understood as a transition from localized to itinerant, where the localized character exists only in the AFM phase and dies out at the QCP, while the itinerant behavior can even coexist in the AFM state.

From the ESR results we observe that there are still some local moments very close to the QCP in the AFM state ($x=0.5$) and none in its vicinity on the nonmagnetic side ($x=0.7$) of the phase diagram. However, from our data, it is difficult to assert whether the QCP in CeIn$_{3-x}$Sn$_x$ is of the itinerant or the localized scenario, and further ESR experiments in samples with different Sn contents are needed to clarify this issue.

\section{Summary}

In summary, our ESR results microscopically show that for the CeIn$_{3-x}$Sn$_x$ system the AFM end member has only highly screened local moments, whereas for the nonmagnetic samples just itinerant bands are probed. For $x=0.5$, in the vicinity of the QCP, on the AFM side of the magnetic phase diagram, the 4$f$ electrons has a dual character, being at the same time localized and itinerant, giving rise to multiband effects.

\begin{acknowledgments}
We thank J. C. B. Monteiro and F. C. G. Gandra for their help with the polycrystalline samples. This work was supported by FAPESP (Grant Nos. 06/55347-1, 06/60440-0, 07/50986-0, and 11/01564-0); CNPq, FINEP, and CAPES (Brazil) and NSF (Grant No. NSF-DMR-0801253) (USA).
\end{acknowledgments}


\begin{thebibliography}{99}
\bibitem{RevHFStewart} G. R. Stewart, Rev. Mod. Phys. \textbf{56}, 755 (1984).
\bibitem{RevLohneysen} H. v. L\"{o}hneysen, A. Rosch, M. Vojta, and P. W\"{o}lfle, Rev. Mod. Phys. \textbf{79}, 1015 (2007).
\bibitem{twofluid_pascoal} S. Nakatsuji, S. Yeo, L. Balicas, Z. Fisk, P. Schlottmann, P. G. Pagliuso, N. O. Moreno, J. L. Sarrao, and J. D. Thompson, Phys. Rev. Lett. \textbf{89}, 106402 (2002).
\bibitem{CeInSn_Lawrence} J.M. Lawrence, Phys. Rev. B \textbf{20}, 3770 (1979).
\bibitem{CeInSn_Euro} P. Pedrazzini, M. G. Berisso, N. Caroca-Canales, M. Deppe, C. Geibel, and J. G. Sereni, Eur. Phys. J. B \textbf{38}, 445 (2004).
\bibitem{CeIn3_SC} N. D. Mathur, F. M. Grosche, S. R. Julian, I. R. Walker, D. M. Freye, R. K. W. Haselwimmer, and G. G. Lonzarich, Nature \textbf{394}, 39 (1998).
\bibitem{CeIn3Cd} N. Berry, E. M. Bittar, C. Capan, P. G. Pagliuso, and Z. Fisk, Phys. Rev. B \textbf{81}, 174413 (2010).
\bibitem{CeIn3_Gru} R. K\"{u}chler, P. Gegenwart, J. Custers, O. Stockert, N. Caroca-Canales, C. Geibel, J. G. Sereni, and F. Steglich, Phys. Rev. Lett. \textbf{96}, 256403 (2006).
\bibitem{CeIn3Sn025} A. V. Silhanek, T. Ebihara, N. Harrison, M. Jaime, K. Tezuka, V. Fanelli, and C. D. Batista, Phys. Rev. Lett. \textbf{96}, 206401 (2006).
\bibitem{dHvA_CeIn3} R. Settai \textit{et al.}, J. Phys. Soc. Jpn. \textbf{74}, 3016 (2005).
\bibitem{NQRCep} S. Kawasaki, M. Yashima, Y. Kitaoka, K. Takeda, K. Shimizu, Y. Oishi, M. Takata, T. C. Kobayashi, H. Harima, S. Araki, H. Shishido, R. Settai, and Y. \={O}nuki, Phys. Rev. B \textbf{77}, 064508 (2008).
\bibitem{ESR_La} E. M. Bittar, C. Adriano, C. Giles, C. Rettori, Z. Fisk, and P. G. Pagliuso, J. Phys.: Condens. Matter \textbf{23}, 455701 (2011).
\bibitem{gambke} T. Gambke, B. Elschner, and L. L. Hirst, Phys. Rev. Lett. \textbf{40}, 1290 (1978).
\bibitem{Venegas} J. G. S. Duque, R. R. Urbano, P. A. Venegas, P. G. Pagliuso, C. Rettori, Z. Fisk, and S. B. Oseroff, Phys. Rev. B \textbf{76}, 125114 (2007) and references therein.
\bibitem{ESR_CeIn3} E. M. Bittar, J. G. S. Duque, P. A. Venegas, C. Rettori, and P. G. Pagliuso, Physica B \textbf{404}, 2995 (2009).
\bibitem{LaInSnrho} W. D. Grobman, Phys. Rev. B \textbf{5}, 2924 (1972).
\bibitem{ref20}  C. Rettori, H. M. Kim, E. P. Chock, and D. Davidov, Phys. Rev. B \textbf{10}, 1826 (1974).
\bibitem{EPR_barnes} S. E. Barnes, Adv. Phys. \textbf{30}, 801 (1981).
\bibitem{DavidovSSC12}  D. Davidov, K. Maki, R. Orbach, C. Rettori, and E.P. Chock, Solid State Commun. \textbf{12}, 621 (1973).
\bibitem{Gaston} G. E. Barberis, D. Davidov, J. P. Donoso, C. Rettori, J. F. Suassuna, and H. D. Dokter, Phys. Rev. B \textbf{19}, 5495 (1979); A. Troper and A. A. Gomes, \textit{ibid.} \textbf{34}, 6487 (1986).
\bibitem{Yafet} Y. Yafet and V. Jaccarino, Phys. Rev. \textbf{133}, A1630 (1964).
\bibitem{ESRYbAl3} R. R. Urbano, E. M. Bittar, M. A. Pires, L. Mendonça Ferreira, L. Bufaiçal, C. Rettori, P. G. Pagliuso, B. Magill, S. B. Oseroff, J. D. Thompson, and J. L. Sarrao, Phys. Rev. B \textbf{75}, 045107 (2007).
\bibitem{Pinto} J. W. M. Pinto and H. O. Frota, Phys. Rev. B \textbf{64}, 092404 (2001); A. Ghosh, M. S. Gusm\~{a}o, and H. O. Frota, Eur. Phys. J. B \textit{in press}.
\bibitem{ESR_Pd3} T. Gambke and B. Elschner, J. de Phys. Colloques \textbf{40}, C5-331 (1979).
\bibitem{LaInSnSC} A. M. Toxen, R. J. Gambino, and L. B. Welsh, Phys. Rev. B \textbf{8}, 90 (1973).
\bibitem{Ani_CeIn1Sn2} A. P. Murani, A. Severing, M. Enderle, P. Steffens, and D. Richard, Phys. Rev. Lett. \textbf{101}, 206405 (2008).
\end{thebibliography}
\end{document}